\documentclass{elsart}

\usepackage{graphics}
\usepackage{graphicx}

\usepackage{amssymb}

\begin{document}

\begin{frontmatter}



\title{Dendritic side-branching with anisotropic viscous fingering}


\author{Tomokazu Honda, Haruo Honjo, Hiroaki Katsuragi}

\address{Department of Applied Science for Electronics and Materials,  \\ Interdisciplinary Graduate School of Engineering Science, \\ Kyushu University, \\ Kasuga,  Fukuoka 816-8580, Japan }

\begin{abstract}
 We studied dendritic side-branching mechanism in the experiment of anisotropic viscous fingering. We measured the time dependence of growth speed of side-branch and the envelop of side-branches. We found that the speed of side-branch gets to be faster than one of the stem and the growth exponent of the speed changes at a certain time. The envelope of side-branches is represented as $Y\sim X^{1.47\pm 0.08}$. 
\end{abstract}

\end{frontmatter}

\section{Introduction}
 Dendritic patterm formation(dendrite) observed in crystal growth is a non-linear and non-equilibirum phenomenon~\cite{1,2}.
The dendrite has been intensively studied with regard to the tip velocity and the shape near the tip region.
The growth law of the dendrite is expressed as $v\rho ^{2} = const.$, where $v$ is the tip velocity and $\rho$ is the radius of curvature of the parabolic tip~\cite{3,4}.
The parabolic tip is linearly stable and the occurrence of side-branches is considered to result from some noise efects~\cite{5}.
However, the dendrite is characterized by the well-developed side-branches.
The side-branching mechanism at the rear of the tip has not been understood well, where there occurs the growth competition among side-branches.
Recently, Li and Beckermann analyzed the shape of the envelope of  the side-branches for pure succinonitorile under the microgravity, and found that the envelope grows with a power law of exponent $0.859$ as a function of the distance from the tip~\cite{6}.
In addition, Corrigan et al. obtained the the exponent $0.852$ under microgravity and the exponent $0.902$ under terrestrial condition~\cite{7}. 
These exponents are less than $1$, and it implies that the envelope angle decreaes as a function of distance from the tip.
In this paper we present the experimental results of the dendrite not in crystal growth but in two-dimensional viscous fingering
system introducing four-fold symmetry.
Because the viscous dendrite is much larger($cm$ of order) than the crystal dendrite($\mu m$ of order) and we can easily change the growth symmetry of the system.
\section{Experimental setup}
 We used a Hele-Shaw cell($0.08mm$ thick, $1000mm\times 500mm$, Fig.1). The bottom plate of the cell is made of a stainless steel etched with square lattice having four-fold symmetry(Fig. 1).
We used silicon oil(density $1.0 g/cm^{3}$, viscousty; $\mu=50 poise$, and surface tention between the silicon oil and air; $\sigma=20.8 mN/m$) as a high viscous fluid and nitrogen gas as a low viscous fluid.
Nitrogen gas was first kept in a vessel as a constant pressure reservoir and we injected the nitrogen gas into the cell filled with the silicon oil at constant injection pressure $\Delta P$;$\Delta P=[nitrogen gas pressure]-[atmospheric pressure]$.
In each experiment, $\Delta P$ was changed from $8kPa$ to $13kPa$ on the increase fo $1kPa$.
The image of growing dendrites were taken with a CCD camera and recorded on a hard disk in PC. The sequence of images were processed every $0.3 seconds$.
\section{Experimental results}
 Fig. 2 shows a viscous dendrite at $\Delta P=12kPa$. In this patterm, side-branches are well-developed in the region far from the tip .
All branches grow along the groove of fourfold symmetry. Therefore our dendrite is recognized as a kinetic dendrite, since growth directin is determined by the kinetic anisotropy~\cite{8}.
The stem speed; $V_{stem}(t)$, is constant($56.8\pm0.5mm/sec.$).
We measured the time dependence of side-branch's tip(Fig.3) and the speed of side-branch; $V_{y}(t)$(Fig. 4), which keeps growing screeing the neighboring side-branches.
Fig. 4 shows the time dependence of speed of the side-branch(pointed out with a white arrow in Fig. 2), and we can confirm that the side-branch acceleratory grows and the speed gets to be faster than one of the stem after $3.3 seconds$.
Fig. 5 shows log-log plot of $V_{y}(t)$ of the side-branch, which chages the growth exponent at a cetain time$\tau$($=2.5sec.$).
In Fig. 6, we show log-log plot of the time dependence of side-branch's speed varying $\Delta P$, where the speed and the time are normalized with each stem speed $V_{stem}$ and $\tau$, respectively. 
The speed of all side-branches changes the growth exponent at a cetain time $\tau$, and the values of $\tau$, do not almost depend on $\Delta P$($\tau\sim2.5sec.$). The average exponent before $\tau$ is $0.51\pm0.02$ and the average exponent after $\tau$ is $1.39\pm0.05$.
In this experiment, an envelope of side-branches is defined as the curve connected all tips of surviving side-branches(Fig. 7).
We have found that the envelope at $\Delta P=12kPa$ at various time has a fitting curve; $Y\sim X^{1.36\pm0.02}$(red line in Fig. 8).
And Fig. 8 also shows the parabolic surface fitting cure;$Y\sim X^{0.5}$(dashedline). The envelopes for the other $\Delta P$ show the same behavior and the average exponent of envelope is $1.47\pm 0.08$.
\section{Discussion }
 We have performed an experiment of kinetic dendrite with anisotropic viscous fingering. Studying the envelope near the tip for pure succinonitrole dendrite under microgravity, Q. Li and C. Beckermann obtained the exponent of the envelope $0.859$~\cite{6}, and D.P.Corrigan et al. obtained the exponent of the envelope $0.852$ under microgravity and $0.902$ under terrestrial condition~\cite{7}. 
However, our results are $0.5$ and $1.47\pm 0.08$. The exponent $0.5$ means a parabola and this value should be larger as mentioned above, and this discrepancy results from our less measured resolution.
 The global dendrite is shown with well-developed side-branches, and the exponent of the envelope is $1.47\pm 0.08$. 
 Then we can image that a dendritic envelope has three regions; a parabolic tip region, the region formed by premature side-branches, and the region formed by welldeveloped side-branches. 
 We emphasize that the distance between side-branches ($66.7mm$) is shorter than the diffusion length ($130mm$) in the region of the exponent $1.36$.
 And the side-branches affect each other (growth competition). The exponent is a similar value to the slope ($1.39$) in Fig. 6. 
 This result may not be trivial because the envelope consists of many well-developed side-branches that do not grow with the same interval. 
 After the growth competition, which means that the side-branches spacing is larger than the diffusion length, we conjecture that the faster side-branch could grow independently with the same speed as the stem, and we need much larger cell to study such growth.
The reason why the growth exponent changes at $\tau$ is as follows.
The diffusion field(pressure field) tends to concentrate to premature side-branches with nearly equal strength.
However, the diffusion field concentrates to faster(well-developed) side-branches and makes them accelerate.

\vspace*{1cm}

\begin{figure}

\begin{center}
\includegraphics[width=10cm, clip]{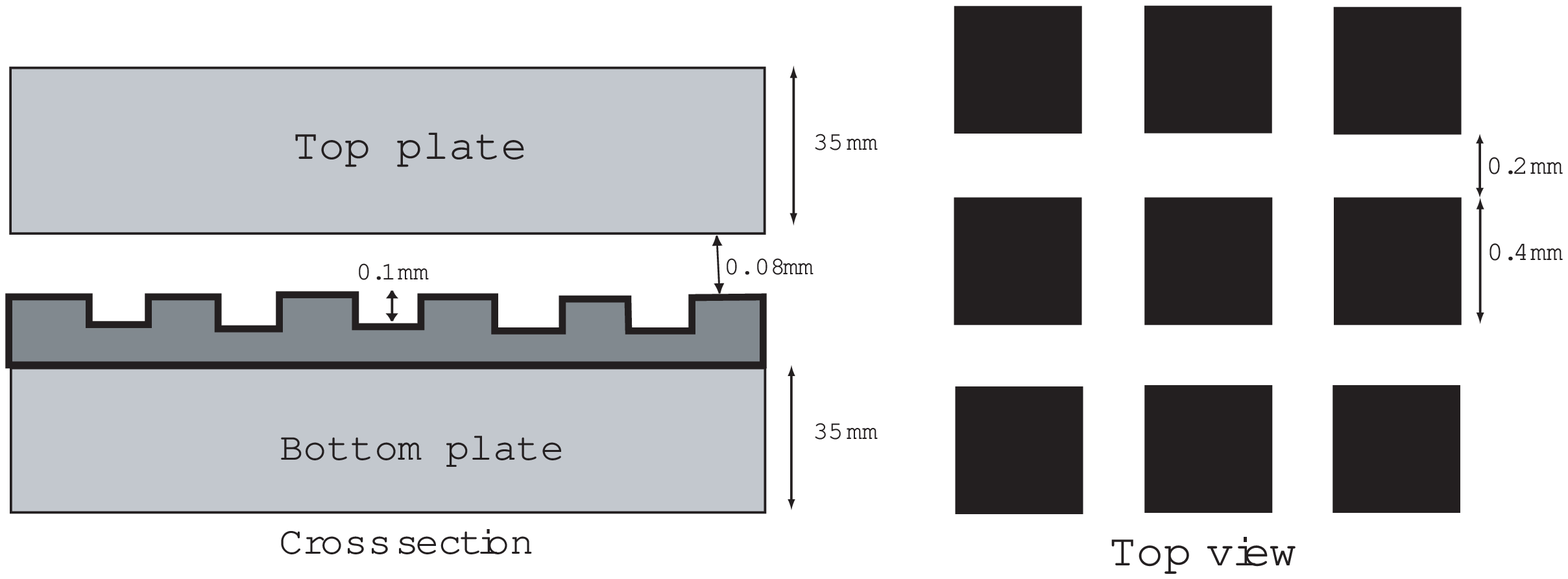}
\caption{Hele-Shaw cell.}
\end{center}

\vspace*{1cm}
\begin{center}
\includegraphics[width=10cm, clip]{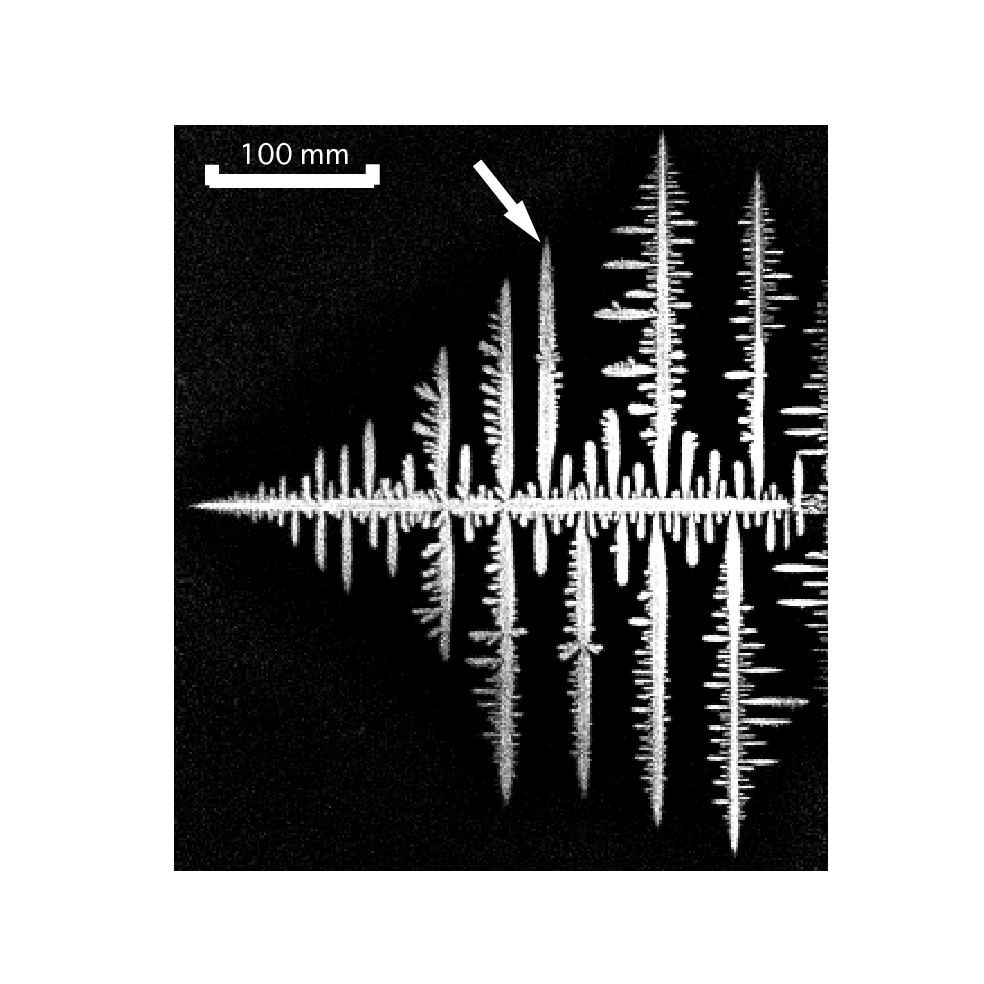}
\caption{Typical viscous dendrite ($\Delta P=12kPa$).}
\end{center}
\end{figure}

\begin{figure}
\begin{center}
\includegraphics[width=10cm, clip]{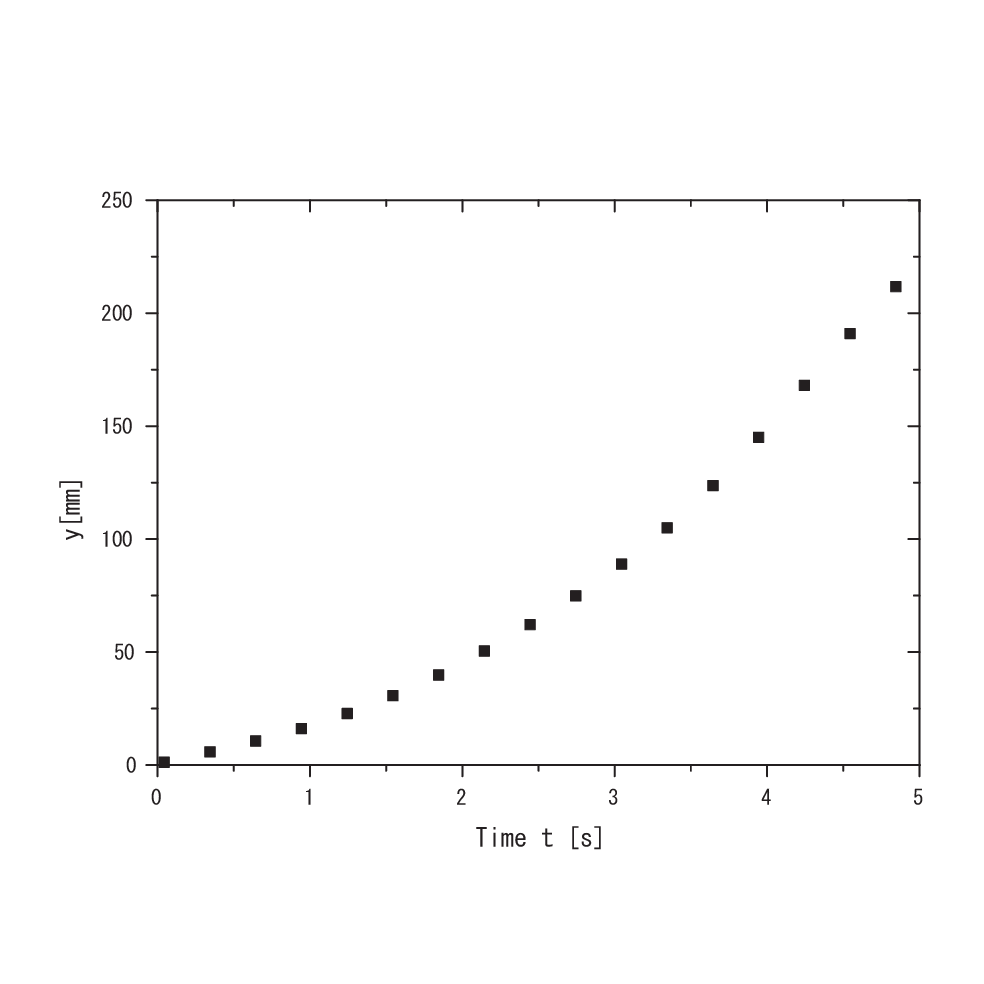}
\caption{Time dependence of side-branch's tip position.}

\end{center}

\begin{center}
\includegraphics[width=10cm, clip]{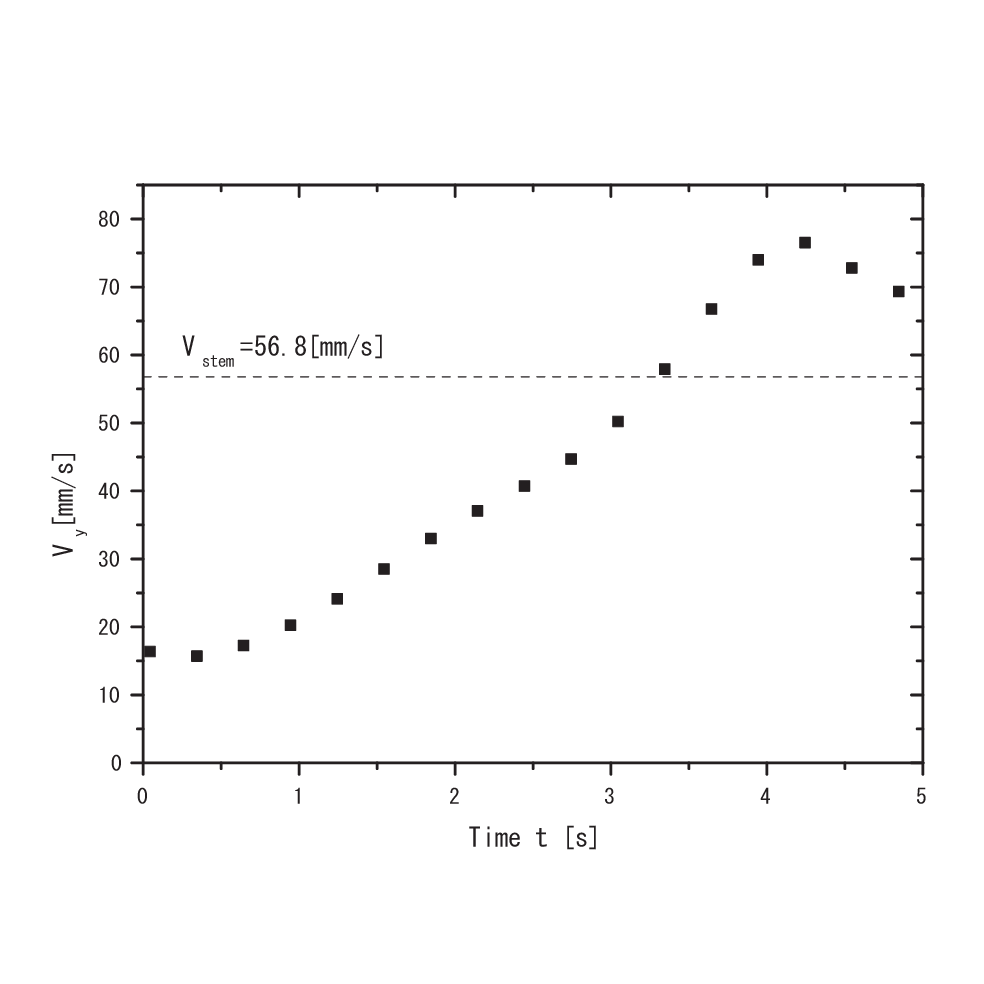}
\caption{Time dependence of side-branch's speed.}
\end{center}
\end{figure}

\begin{figure}
\begin{center}
\includegraphics[width=10cm, clip]{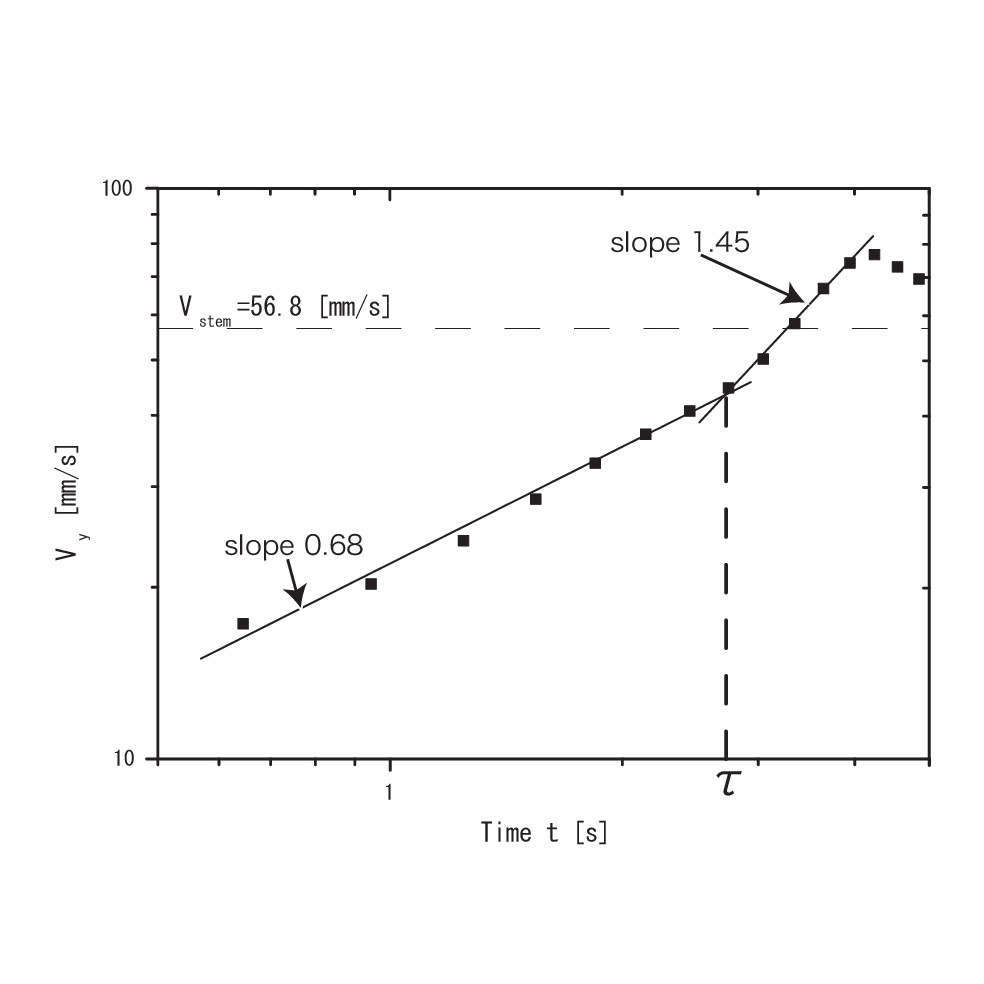}
\caption{Log-log plot of time dependence of side-branch's speed.}
\end{center}

\begin{center}
\includegraphics[width=10cm, clip]{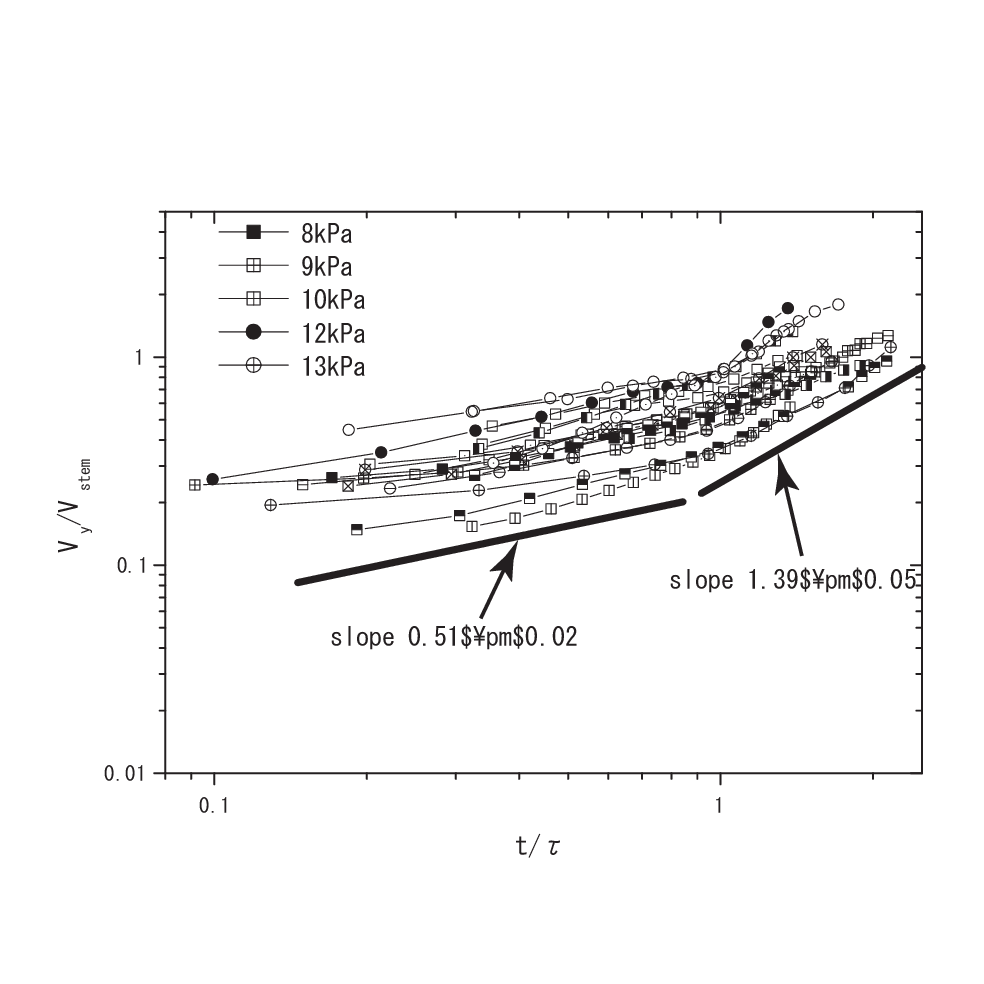}
\caption{Log-log plot of time dependence of speed of all side-branches.}
\end{center}
\end{figure}

\begin{figure}
\begin{center}
\includegraphics[width=10cm, clip]{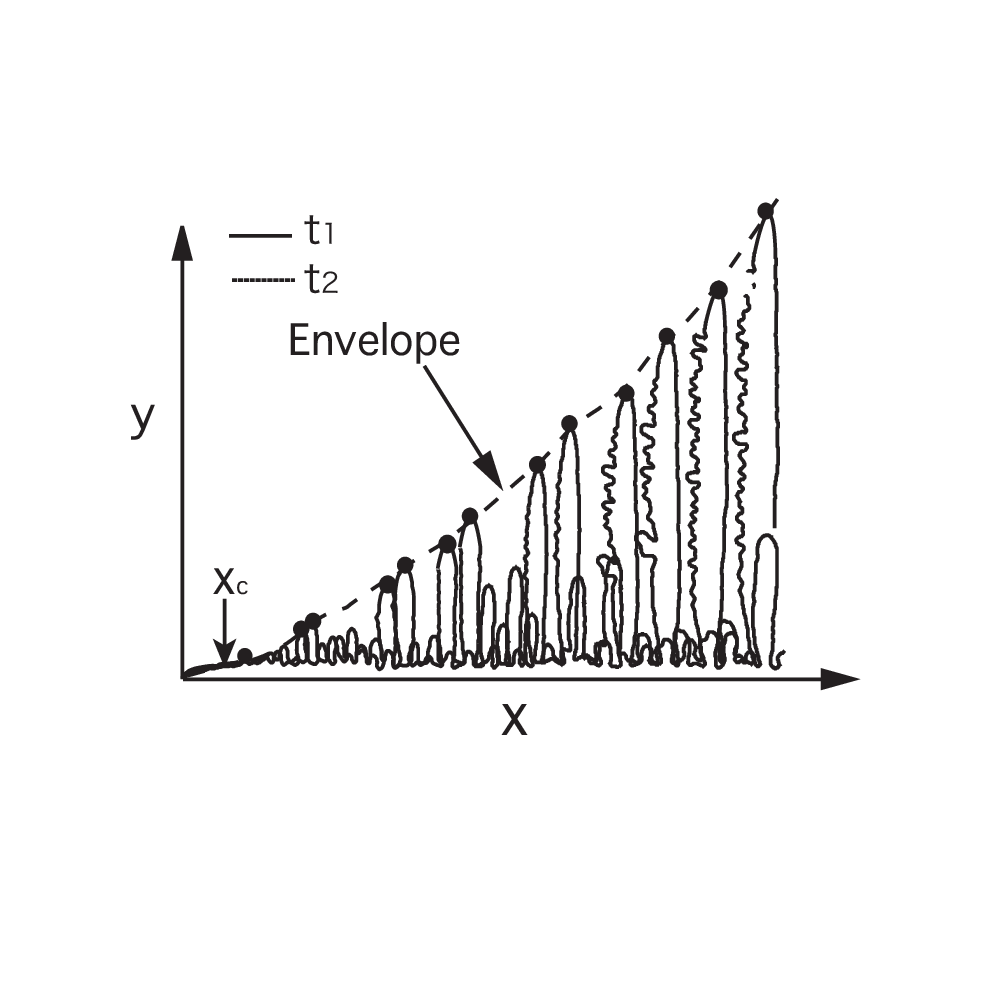}
\caption{Envelope connected all tips of surviving side-branches at different time, $t_{1}$ and $t_{2}$. }
\end{center}

\begin{center}
\includegraphics[width=10cm, clip]{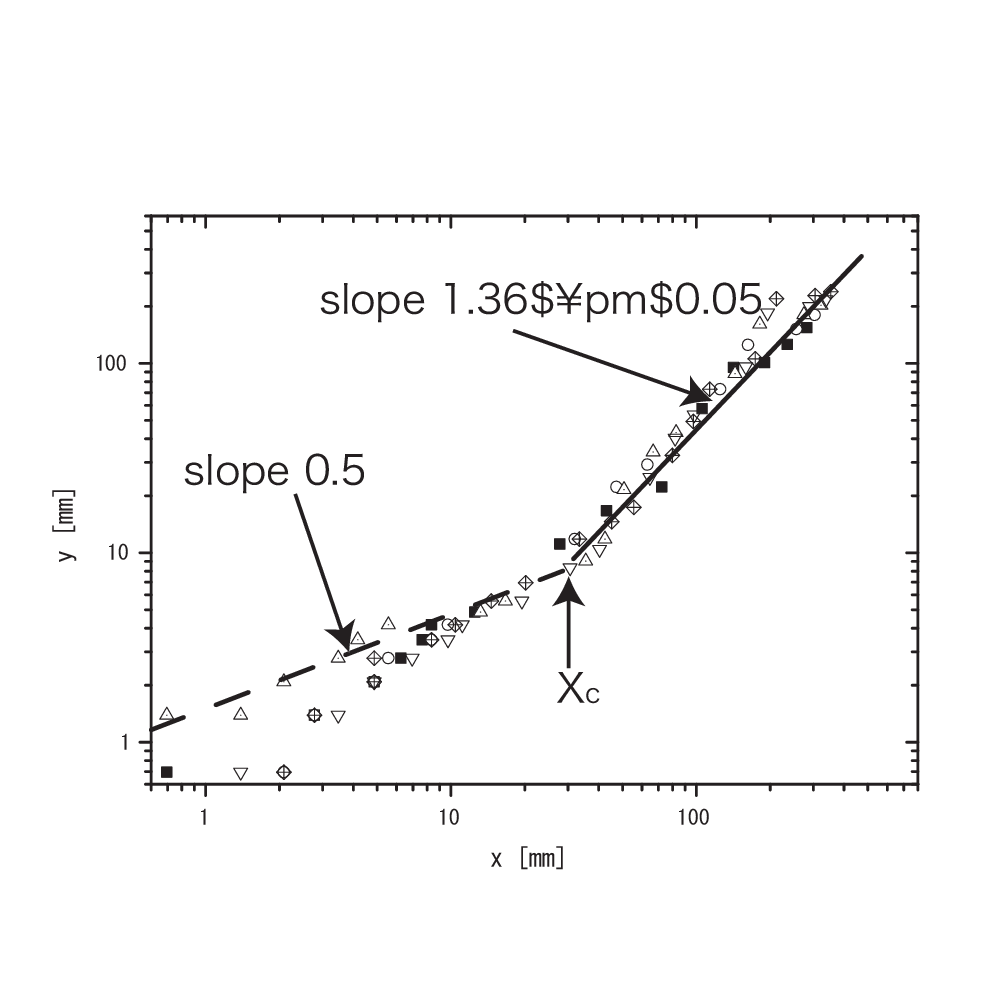}
\caption{Log-log plot of the envelope. The symbols correspond to various times.}
\end{center}

\end{figure}


\begin{thebibliography}{9}

\bibitem{1}
 J. S. Langer,
\textit{Rev. Mod. Phys.} \textbf{52} (1980) 1.

\bibitem{2} 
T. Vicsek,
\textit{Fractal Growth Phenomena (World-Scientific, Singapore, 1991, 2nd ed.)}.

\bibitem{3}
S. C. Huang and M. E. Glicksman,
 \textit{Acta Metal.} \textbf{29} (1981) 701. 

\bibitem{4}
H. Honjo and Y. Sawada, 
\textit{J. Cryst. Growth} \textbf{58} (1982) 297.

\bibitem{5}
A. Dougherty et al., 
\textit{Phys. Rev. Lett.} \textbf{58} (1987) 1652. 

\bibitem{6}
Q. Li and C. Beckermann, 
\textit{Phys. Rev. E} \textbf{57} (1997) 3176. 

\bibitem{7}
D. P. Corrigan et al., 
 \textit{Phys. Rev. E} \textbf{60} (1999) 7217. 

\bibitem{8}
 E. Ben-Jacob et al., 
 \textit{Phys. Rev. A} \textbf{38} (1988) 1370.

\end{thebibliography}
\end{document}